\documentclass[12pt]{article}
\usepackage{epsfig}
\usepackage{amsmath,amsthm, amsfonts}
\usepackage{soul,color}
\usepackage{bm}
\usepackage{hyperref}
\usepackage[backend=biber,natbib=true,style=apa,autocite=inline]{biblatex}
\usepackage{latexsym, graphicx, alltt}
\usepackage{graphicx}
\usepackage{setspace}
\usepackage[table]{xcolor}
\usepackage{float}
\usepackage{booktabs}
\usepackage{doi}
  \usepackage{tabularx}
  \usepackage{multirow}
  \usepackage[title]{appendix}

\usepackage[caption=false]{}
\usepackage{subcaption}
\usepackage{changepage}

\usepackage{framed,varwidth}                             
 \newcount\Comments  
\usepackage{color}
\definecolor{darkgreen}{rgb}{0,0.5,0}
\definecolor{purple}{rgb}{1,0,1}
\newcommand{\kibitz}[2]{\ifnum\Comments=1\textcolor{#1}{#2}\fi}

\renewcommand{\baselinestretch}{1.5}

\begin{document}

\title{Statistical Jump Model for Mixed-Type Data with Missing Data Imputation}

\author{Federico P. Cortese, Antonio Pievatolo\\
{\small Institute for Applied Mathematics and Information Technologies ``E. Magenes'', Milan}\\ 
\small{
National Research Council of Italy} \\
{\small federico.cortese@mi.imati.cnr.it}
}

\maketitle
\date{}
\vspace*{-1cm}
\def\baselinestretch{1.1}
\abstract{In this paper, we address the challenge of clustering mixed-type data with temporal evolution by introducing the statistical jump model for mixed-type data. This novel framework incorporates regime persistence, enhancing interpretability and reducing the frequency of state switches, and efficiently handles missing data. The model is easily interpretable through its state-conditional means and modes, making it accessible to practitioners and policymakers. We validate our approach through extensive simulation studies and an empirical application to air quality data, demonstrating its superiority in inferring persistent air quality regimes compared to the traditional air quality index. Our contributions include a robust method for mixed-type temporal clustering, effective missing data management, and practical insights for environmental monitoring.}

\noindent \vskip2mm \noindent {\sc Keywords: Environmental Monitoring; Mixed-type Data; Missing Data; Regime-switching Models; Unsupervised Learning}
\newpage

\section{Introduction}
\label{sec:intro}

Clustering mixed-type data is particularly challenging when accounting for its temporal evolution \citep{aghabozorgi2015time}. Existing literature on temporal clustering techniques addressed various aspects of this problem, with some exploring the inclusion of temporal persistence \citep{nystrup:2020}. In fact, persistence is crucial for interpretability and practical applications, as frequent state switches can complicate decision-making processes \citep{nystrup2019multi}. 
Moreover, many real-world applications, especially in environmental and ecological studies \citep{kasam2014statistical}, face the issue of missing data, necessitating robust methods to handle such gaps effectively \citep{little2019statistical}.

Regime-switching models \citep{hamilton:1989}, of which hidden Markov models are an example, \citep{bart:farc:penn:13,zucchini:2017, russo2024copula} offer a valuable framework for analyzing time-series data that exhibit different states or regimes over time. These models have been effectively applied in various domains, including financial markets \citep{cortese:2023} and environmental studies \citep{kehagias2004hidden}. The integration of mixed-type data and the handling of missing values in these models still require further development.

The main purpose of this study is to provide a unified framework for clustering mixed-type temporal data, which accounts for regime persistence and handles missing values efficiently. Our proposed method, which we call \textit{statistical jump model for mixed-type data} (JM-mix), enhances interpretability and practical applicability by reducing the frequency of state switches and providing a robust solution for missing data.
Our model is inspired by the statistical jump model of \cite{bemporad:2018}, which \cite{nystrup:2021} extended to address data sparsity, and \cite{aydinhan2024identifying} further adapted for soft clustering.

Our approach is motivated by the need for an advanced yet interpretable clustering method that outperforms existing models such as spectral clustering \citep{von2007tutorial} and $k$-prototypes \citep{huang1998extensions}. Unlike more sophisticated methods like artificial neural networks \citep{liao2007artificial}, our model is straightforward to understand and interpret, making it accessible to practitioners and policymakers.

We make three main contributions to the existing literature. First, we introduce a new method for clustering mixed-type data over time, incorporating persistence in regimes to enhance interpretability and utility. Second, our approach includes a fast and efficient solution for managing missing data, ensuring robust clustering results. Third, we demonstrate the applicability of our method through an empirical study on air quality data. Our results show that our method can infer more persistent air quality regimes compared to the traditional air quality index \citep[AQI,][]{rule2005environmental}.

The rest of the paper is organized as follows. Section \ref{sec:model} introduces the model formulation, estimation procedures, and provides guidance on hyperparameter selection. Section \ref{sec:simstud} presents an extensive simulation study, demonstrating that our proposal outperforms competitive models in terms of clustering accuracy. Section \ref{sec:empstud} showcases an application to air quality data, where the level of air quality is treated as a latent process inferred from observable variables such as weather conditions and pollutant levels. Section \ref{sec:discussion} concludes the paper with a discussion of the findings and implications for future research.

\section{Methodology}
\label{sec:model}

Let us consider data \(\boldsymbol{y} 
\), with $T$ rows (times) and $P$ columns (features).
%
We denote the vector of observations at time $t=1,\ldots,T$, as
$\boldsymbol{y}_{t,\cdot}=(y_{t,1},\ldots,y_{t,P})^\prime\in \mathbb{R}^P$, 
and the vector of observations for feature $p$, $p=1,\ldots,P$, as
$\boldsymbol{y}_{\cdot,p}=(y_{1,p},\ldots,y_{T,p})\in \mathbb{R}^T$. 
Among these 
$P$ features, 
some are continuous, while others are categorical.

We further assume that some observations are missing. We denote a missing observation for feature \( p \) at time \(t\) as \( y^{(miss)}_{t,p} \). When we need to specify an observed value at time $t$, or the vector of observed values for feature $p$, we denote them as \( y^{(obs)}_{t,p} \) and \( \boldsymbol{y}^{(obs)}_{\cdot,p} \), respectively. For simplicity, we may sometimes omit these superscripts.

Regarding the model formulation, we assume that $\boldsymbol{\mu}_k=(\mu_{k,1},\ldots,\mu_{k,P})^\prime$, $k=1,\ldots,K$, consists of the state conditional means for continuous 
variables, and of the state conditional modes for categorical variables \citep{huang1998extensions}.

A statistical jump model \citep{bemporad:2018} transitions among a finite set of states or submodels to appropriately represent a sequence of data, while  considering the order of the data points. Similar to \cite{nystrup:2020}, we propose fitting a JM-mix 
with $K$ states by minimizing the following objective function
\begin{equation}
    \label{eq:mixedJM}
    \sum_{t=1}^{T-1}\left[g(\boldsymbol{y}_{ t,\cdot},\boldsymbol{\mu}_{s_{ t}})+\lambda \mathbb{I}(s_{ t}\neq s_{{ {t+1}}})\right]+g(\boldsymbol{y}_{ T,\cdot},\boldsymbol{\mu}_{s_{ T}}),
\end{equation}
%
over the latent state sequence $\boldsymbol{s}=(s_{ 1},\ldots,s_{ T})^\prime$ and the model parameters $\boldsymbol{\mu}_k=(\mu_{k,1},\ldots,\mu_{k,P})^\prime$.
$g(\cdot,\cdot)$ is the Gower distance \citep{gower1971general,tuerhong2014gower} (Equation \eqref{eq:gower} of Appendix \ref{appendix A}), which is a metric used to measure dissimilarity between mixed-type variables, and $\lambda \geq 0$ is a temporal jump penalty as in \cite{nystrup:2020} and \cite{nystrup:2021}. It balances data fitting with state sequence stability, where a higher $\lambda$ value makes the model more likely to stay in the same state, thus reducing the frequency of state changes.

\subsection{Model estimation}

Following \cite{nystrup:2021}, jump models can be efficiently fitted using a coordinate descent algorithm, which iteratively alternates between optimizing model parameters for a fixed state sequence, and updating the state sequence, based on these parameters, to minimize the objective function \eqref{eq:mixedJM} \citep{bemporad:2018}. This iterative process is repeated 10 times or concludes earlier if the state sequence stabilizes. Achieving the global optimum is not guaranteed due to the dependency on the initial state sequence: to enhance solution quality, the algorithm is executed multiple times with various initial state sequences in parallel, selecting the model with the lowest objective value \citep{arthur2007k}. 
The most likely sequence of states is found using dynamic programming: in particular, we use a reversed Viterbi algorithm \citep{viterbi:1967} (see \cite{nystrup:2020} for details).

We handle missing data by imputing the missing values at each iteration of the estimation process with the corresponding estimates of state-conditional means or modes, depending on the type of data. This approach is a modification of the $k$-pods method from \cite{chi2016k}, which is proven to be effective even when the missingness mechanism is unknown, external information is unavailable, and there is significant missingness in the data. For alternative imputation strategies when clustering incomplete mixed-type data, please refer to \cite{aschenbruck2023imputation}.

The estimation process can be summarized as follows.
\begin{itemize}
    \item[(a)] Initialize state sequence $s_{ 1},\ldots,s_{ T}$.
    \item[(b)] Initialize missing values with mean or mode of each feature, i.e. 
    $y^{(miss)}_{{ t},p}=\boldsymbol{\bar{y}}^{(obs)}_{\cdot, p}$ if $\boldsymbol{y}_{\cdot, p}$ is continuous and $y^{(miss)}_{t,p}=\operatorname{Mode}\left(\boldsymbol{y}^{(obs)}_{\cdot, p}\right)$ if $\boldsymbol{y}_{\cdot, p}$ is categorical.
    \item[(c)] Iterate the following steps for $j \in 1,\ldots$, until $\boldsymbol{s}^j=\boldsymbol{s}^{j-1}$:
    \begin{itemize}
        \item[(i)] Fit model parameters $\boldsymbol{\mu}=\{\boldsymbol{\mu}_1,\ldots,\boldsymbol{\mu}_K\}$
        $$\boldsymbol{\mu}^j=\operatorname*{argmin}_{\boldsymbol{\mu}}\sum_{t=1}^T g(\boldsymbol{y}_{ t,\cdot},\boldsymbol{\mu}_{s_{{ t}}^{j-1}}).$$
         \item[(ii)]
         Update missing values
         $$y^{(miss)}_{{ t},p}=\mu^j_{s_{ t}, p}, \,\, t=1,\ldots,T, \, p=1,\ldots,P,$$
         where $\mu^j_{s_{ t}, p}$ is the mean (or mode) of feature $p$ in cluster $s_{ t}$. For clarity, the superscript $j-1$ on $s_{ t}$ is omitted.
         
        \item[(iii)]
       Fit state sequence
        $$\boldsymbol{s}^{j}=\operatorname*{argmin}_{\boldsymbol{s}}
        \left\{
         \sum_{t=1}^{T-1} 
         \left[
         g(\boldsymbol{y}_{ t,\cdot},\boldsymbol{\mu}^j_{s_{ t}})+\lambda\mathbb{I}(s_{t}\neq s_{ {t+1}})\right]+g(\boldsymbol{y}_{ T,\cdot},\boldsymbol{\mu}^j_{s_{ T}}) 
        \right\}.
        $$
    \end{itemize}
\end{itemize}


\subsection{Hyperparameters selection}
\label{subsec:hyperpar}
\citet{witten2010framework} proposed selecting hyperparameters for a sparse $K$-means model with the value that maximizes the gap between the between-cluster sum of squares for the actual data and that for random permutations of the data.
Alternatively, one can select hyperparameters based on the specific application, using methods such as 
backtesting. For instance, if the estimated states are used in an investment strategy, \(\lambda\) can be chosen to maximize risk-adjusted returns, accounting for transaction costs \citep{nystrup2019multi,nystrup:2021}. 
Alternatively, if ground truth data is available, accuracy can be maximized against this benchmark.

In the empirical analysis of Section \ref{sec:empstud}, we adopt an information criterion to balance the in-sample goodness of fit with model complexity, thereby avoiding overfitting.
In particular, we use a version of the generalized information criterion (GIC) from \cite{cortese2024generalized}, shown to be effective in selecting the correct hyperparameter values through extensive simulation studies for continuous data. 
The criterion is given by
\begin{align}
    \label{eq:SJMGIC}
\text{GIC} &:= \frac{1}{T} \left\{  (\text{BCD}^S-\text{BCD}) + a_T M\right\} + 2 \left(C (K^S) - C (K)\right),
\end{align}
where \( C(K) := -\log(K) - \frac{P}{2}\log(2\pi) \), and
\[
\text{BCD} = \sum_{k=1}^K  g\left( \boldsymbol{\mu}_k, \bar{\boldsymbol{\mu}} \right)n_k,
\]
is the between cluster total deviance \citep{perssonrobust2023}, essentially the JM-mix version of the between clusters sum of squares \citep{witten2010framework}. \( n_k \) is the number of points in cluster \( k \), and \( \bar{\boldsymbol{\mu}} \) is the vector of unconditional means and modes. $a_T$ 
depends
on the sample size $T$, and possibly the number of parameters and/or features considered; following \cite{fan2013tuning}, we set $a_T=\log(\log(T)) \log(P)$. 
Finally, we set model complexity $M= K(P+\Gamma_\lambda)$,
where \( \Gamma_\lambda = \sum_t \mathbb{I}_{s_t \neq s_{t-1}} \) represents the number of jumps across states, depending indirectly on the hyperparameter \( \lambda \); as \( \lambda \) increases, \( \Gamma_\lambda \) decreases.

In equation \eqref{eq:SJMGIC}, the superscript \( S \) denotes the saturated model, which is a JM-mix without any penalty on jumps (i.e., \( \lambda^S = 0 \)) and with the maximum number of states set at \( K^S = 6 \).\footnote{The selection of the number of states for the saturated model, \( K^S \), is less straightforward than that of \( \lambda^S \). For details on this, please refer to \cite{cortese2023b} or \cite{cortese2024generalized}.}

\section{Simulation study}
\label{sec:simstud}
In this section, we show a series of simulation studies assessing the JM-mix performance under various scenarios. After outlining the simulation setup in Subsection \ref{subsec:simsetup}, Subsection \ref{subsec:cpmpleteequally} investigates performance of the model with complete data, while Subsection \ref{subsec:uncomplete} focuses on incomplete data.

\subsection{Simulation setup}
\label{subsec:simsetup}

We simulate features from a multivariate Gaussian HMM with $3$ latent states,
\begin{equation}
    \label{eq:modform}
    \boldsymbol{y}_{t,\cdot} \mid s_{t} \sim N_P(\boldsymbol{\mu}_{s_{t}},  \boldsymbol{\Sigma}_P),
\end{equation}
where \(\boldsymbol{\mu}_{s_{t}} = (\mu_k, \ldots, \mu_k)^\prime\in \mathbb{R}^P\), with \(\mu_1 = \mu\), \(\mu_2 = 0\), and \(\mu_3 = -\mu\).  
The state-conditional covariance matrix $\boldsymbol{\Sigma}_P$ has diagonal elements equal to 1 and off-diagonal elements $\rho_{ij}=\rho$, $i,j=1,\ldots,P$, $i\neq j$.
We consider three possible simulation setups for the parameters of the data generating process in \eqref{eq:modform}, namely
\begin{enumerate}
    \item[(1)] \(\mu = 1\), \(\rho = 0\);
    \item[(2)] \(\mu = 1\), \(\rho = 0.20\);
    \item[(3)] \(\mu = 0.50\), \(\rho = 0\).
\end{enumerate}

The latent process $\{s_t\}$ is a Markov chain of first order with $3$ states, with initial probabilities $\pi_i=1/3$, $\forall i=1,\ldots,3$, and transition probability matrix denoted by $\boldsymbol{\Pi}\in \mathbb{R}^{
3\times 3
}$, with elements $\pi_{ij}$, $i,j=1,\ldots,3$. 
We simulate latent states 
considering self-transition probabilities \( \pi_{kk} = 0.95, \, k=1,2,3 \), and probabilities of transitioning to a different state $\pi_{ij}=(1-\pi_{kk})/2$, $i\neq j$. 

Next, we transform half of these features (\(P/2\))  into categorical variables with 3 levels \( l = 1, 2, 3 \), with probabilities 
$$
\mathbb{P}(l = \tilde{k} \mid s_{t} = k) = 
\begin{cases} 
0.80 & \text{if } \tilde{k} = k, \\
0.10 & \text{if } \tilde{k} \neq k ,
\end{cases}
$$
for \( \tilde{k} = 1, 2, 3 \). 

We vary \( T \in \{50, 100, 500\} \) and \( P \in \{25, 50, 75\} \), and simulate 100 datasets as described above for each combination of \( T \) and \( P \).

For the JM-mix, for each combination and each simulated dataset, we vary \(\lambda\) in a grid between 0 and 1 with a step size of 0.05, and select the optimal \(\lambda\) as the value that maximizes the adjusted rand index \citep[][ARI,]{hubert:1985} computed between true and estimated states sequences. See Appendix \ref{appendix B} for details on how to compute this index. 

In the following simulation studies and the empirical application in Section \ref{sec:empstud}, 
we observe that upper bounds for $\lambda$ greater than approximately 0.9 yield stable results for the decoded state sequence and estimated parameters;
consequently, we search for $\lambda$ in $(0,1)$.

We compare our proposal to two clustering alternatives: spectral clustering \citep[spClust,][]{jia2014latest,von2007tutorial} and a modified version of the $k$-means method that incorporates continuous and categorical variables by \cite{huang1998extensions}, known as $k$-prototypes ($k$-prot). For the latter, we consider a version that considers the Gower distance metric, as described by \cite{szepannek2024clustering}.

spClust involves constructing a similarity graph from the data points, computing the Laplacian of the graph, and then performing dimensionality reduction before applying a standard clustering algorithm like $k$-means. This approach can capture complex relationships in the data that $k$-means might miss.
$k$-prot is essentially our proposal when no penalty on jumps is applied (i.e., JM-mix with $\lambda = 0$).

\subsection{Performance evaluation with complete data}
\label{subsec:cpmpleteequally}

Table \ref{tab:combinedSetups} illustrates the ARI computed between true and estimated sequences of states, for each simulated setup, using the three clustering models, and across different sample sizes.

The JM-mix consistently outperforms $k$-prot and spClust, demonstrating its robustness in various contexts. When \(\mu=1\) and $\rho=0$ (setup 1), JM-mix notably excels, achieving ARIs up to 0.99.

As correlation is introduced in setup 2 (\(\rho=0.2\)), JM-mix continues to show superior performance, particularly at larger sample sizes, indicating its effectiveness in handling both correlation and larger datasets. For instance, at \(T=500\), JM-mix achieves ARIs close to 0.92, outperforming the other two models. The lower performance in setup 2 is probably due to the presence of feature correlation that none of the methods explicitly model. Interestingly, spClust, which should theoretically capture complex relationships, performs worse than $k$-prot in this scenario.

Setup 3 explores a smaller mean (\(\mu=0.5\)) without correlation. Here, all models improve their performance compared to setup 2, with JM-mix still leading, especially at \(T=100\) where it reaches a perfect ARI of 1.00. spClust also shows its best performance in this setup, particularly at larger sample sizes, reaching ARIs up to 0.99, which closely rivals JM-mix.

We attribute the superior performance of JM-mix to its consideration of the temporal nature of the data, which other methods do not account for. Additionally, as the number of features increases, the performance of all methods improves, likely due to the increased information about the latent state provided by a larger number of observations that share the same conditional distribution.
\begin{table}[h]
\centering
\caption{
Adjusted Rand Index (ARI) computed between true and estimated latent sequences for the three models: $k$-prototypes ($k$-prot), spectral clustering (spClust), and jump model for mixed data (JM-mix) for the three setups. Monte Carlo standard deviations are in parentheses.
}
\label{tab:combinedSetups}
\begin{tabular}{ccccc}
\toprule
\multicolumn{5}{c}{Setup 1} \\
\midrule
$T$ & Method & $P=25$ & $P=50$ & $P=75$ \\
\midrule
\multirow{3}{*}{50} & $k$-prot & 0.35 (0.27) & 0.52 (0.27) & 0.60 (0.28) \\
                    & spClust & 0.54 (0.34) & 0.66 (0.33) & 0.79 (0.35) \\
                    & JM-mix & \textbf{0.82} (0.27) & \textbf{0.83} (0.26) & \textbf{0.89} (0.23) \\
\midrule
\multirow{3}{*}{100} & $k$-prot & 0.46 (0.18) & 0.60 (0.20) & 0.79 (0.22) \\
                     & spClust & 0.41 (0.16) & 0.78 (0.20) & 0.82 (0.19) \\
                     & JM-mix & \textbf{0.80} (0.17) & \textbf{0.93} (0.12) & \textbf{0.94} (0.10) \\
\midrule
\multirow{3}{*}{500} & $k$-prot & 0.41 (0.11) & 0.94 (0.15) & \textbf{0.99} (0.03) \\
                     & spClust & 0.78 (0.08) & 0.93 (0.05) & 0.98 (0.03) \\
                     & JM-mix & \textbf{0.97} (0.08) & \textbf{0.97} (0.02) & \textbf{0.99} (0.02) \\
                     \toprule
\multicolumn{5}{c}{Setup 2} \\
\midrule
$T$ & Method & $P=25$ & $P=50$ & $P=75$ \\
\midrule
\multirow{3}{*}{50} & $k$-prot & 0.35 (0.25) & 0.39 (0.23) & 0.46 (0.23) \\
                    & spClust & 0.26 (0.25) & 0.35 (0.23) & 0.35 (0.23) \\
                    & JM-mix & \textbf{0.70} (0.25) & \textbf{0.82} (0.23) & \textbf{0.81} (0.23) \\
\midrule
\multirow{3}{*}{100} & $k$-prot & 0.40 (0.18) & 0.47 (0.17) & 0.48 (0.18) \\
                     & spClust & 0.34 (0.17) & 0.42 (0.14) & 0.41 (0.14) \\
                     & JM-mix & \textbf{0.73} (0.17) & \textbf{0.78} (0.17) & \textbf{0.79} (0.18) \\
\midrule
\multirow{3}{*}{500} & $k$-prot & 0.38 (0.10) & 0.45 (0.09) & 0.49 (0.08) \\
                     & spClust & 0.12 (0.16) & 0.44 (0.10) & 0.45 (0.09) \\
                     & JM-mix & \textbf{0.82} (0.10) & \textbf{0.91} (0.09) & \textbf{0.92} (0.08) \\
\toprule
\multicolumn{5}{c}{Setup 3} \\
\midrule
$T$ & Method & $P=25$ & $P=50$ & $P=75$ \\
\midrule
\multirow{3}{*}{50} & $k$-prot & 0.46 (0.26) & 0.51 (0.28) & 0.55 (0.30) \\
                    & spClust & 0.55 (0.30) & 0.82 (0.33) & 0.87 (0.34) \\
                    & JM-mix & \textbf{0.88} (0.26) & \textbf{0.89} (0.22) & \textbf{0.90} (0.21) \\
\midrule
\multirow{3}{*}{100} & $k$-prot & 0.60 (0.15) & 0.74 (0.15) & 0.83 (0.17) \\
                     & spClust & 0.70 (0.19) & 0.90 (0.17) & 0.96 (0.16) \\
                     & JM-mix & \textbf{0.93} (0.14) & \textbf{0.96} (0.13) & \textbf{1.00} (0.12) \\
\midrule
\multirow{3}{*}{500} & $k$-prot & 0.64 (0.06) & 0.75 (0.06) & 0.84 (0.03) \\
                     & spClust & 0.86 (0.05) & \textbf{0.97} (0.02) & \textbf{0.99} (0.01) \\
                     & JM-mix & \textbf{0.92} (0.04) &\textbf{0.97} (0.05) & \textbf{0.99} (0.01) \\
\bottomrule
\end{tabular}
\end{table}


\subsection{Performance evaluation with incomplete data}
\label{subsec:uncomplete}

Following \cite{fu2024filling}, we categorize missing data into two types: random and continuous. Random missing data consists of short, sporadic gaps, while continuous missing data spans over longer periods. Random gaps are typically easier to handle, whereas continuous gaps pose more significant challenges for imputation and predictive modeling.

To explore the effects of both types of missing data on classification performance, we introduce them into the datasets simulated based on setup 1 for all features, accounting for 10\% and 20\% of the total observations. We use different seeds for each simulation to ensure variability and robustness in the ability of our method to handle missing data under varying conditions.
It is noteworthy that spectral clustering was not considered in this analysis as it does not handle missing values.

Results from Tables \ref{tab:contmiss} and \ref{tab:randmiss} indicate that the classification performance of the JM-mix  remains robust despite the presence of missing data, whether continuous or random. This robustness is evident when comparing these results to those in Table \ref{tab:combinedSetups}, where JM-mix consistently outperforms other models with complete data. 

The average imputation error, measured by the Gower distance between true and imputed data, decreases with increasing $T$ in both approaches. As expected, the imputation error is generally higher in the continuous missing data scenario. However, the differences in imputation accuracy are not significant, with JM-mix showing a slight advantage, reducing imputation errors by approximately 0.01.

\begin{table}[h]
\centering
\caption{
Adjusted Rand Index (ARI) 
for $k$-prototypes ($k$-prot) and jump model for mixed data (JM-mix) with 10\% and 20\% continuous missing data (setup 1). 
Monte Carlo standard deviations are in parentheses.
}
\label{tab:contmiss}
\begin{tabular}{ccccc}
\toprule
\multicolumn{5}{c}{10\% Continuous Missing Data} \\
\midrule
$T$ & Method & $P=25$ & $P=50$ & $P=75$ \\
\midrule
\multirow{2}{*}{50} & $k$-prot & 0.28 (0.25) & 0.45 (0.26) & 0.53 (0.28) \\
                    & JM-mix & \textbf{0.71} (0.25) & \textbf{0.84} (0.23) & \textbf{0.89} (0.23) \\
\midrule
\multirow{2}{*}{100} & $k$-prot & 0.43 (0.18) & 0.53 (0.19) & 0.72 (0.18) \\
                     & JM-mix & \textbf{0.70} (0.17) & \textbf{0.88} (0.17) & \textbf{0.90} (0.18) \\
\midrule
\multirow{2}{*}{500} & $k$-prot & 0.40 (0.13) & 0.86 (0.12) & 0.94 (0.05) \\
                     & JM-mix & \textbf{0.91} (0.12) & \textbf{0.95} (0.06) & \textbf{0.96} (0.05) \\
\bottomrule
\toprule
\multicolumn{5}{c}{20\% Continuous Missing Data} \\
\midrule
$T$ & Method & $P=25$ & $P=50$ & $P=75$ \\
\midrule
\multirow{2}{*}{50} & $k$-prot & 0.21 (0.26) & 0.40 (0.25) & 0.49 (0.26) \\
                    & JM-mix & \textbf{0.62} (0.26) & \textbf{0.77} (0.25) & \textbf{0.82} (0.26) \\
\midrule
\multirow{2}{*}{100} & $k$-prot & 0.36 (0.21) & 0.48 (0.18) & 0.56 (0.16) \\
                     & JM-mix & \textbf{0.67} (0.21) & \textbf{0.78} (0.18) & \textbf{0.79} (0.16) \\
\midrule
\multirow{2}{*}{500} & $k$-prot & 0.35 (0.12) & 0.61 (0.13) & 0.70 (0.10) \\
                     & JM-mix & \textbf{0.70} (0.12) & \textbf{0.84} (0.13) & \textbf{0.82} (0.10) \\
\bottomrule
\end{tabular}
\end{table}
\begin{table}[h]
\centering
\caption{
Adjusted Rand Index (ARI) 
for $k$-prototypes ($k$-prot) and jump model for mixed data (JM-mix) with 10\% and 20\% random missing data (setup 1). 
Monte Carlo standard deviations are in parentheses.
}
\label{tab:randmiss}
\begin{tabular}{ccccc}
\toprule
\multicolumn{5}{c}{10\% Random Missing Data} \\
\midrule
$T$ & Method & $P=25$ & $P=50$ & $P=75$ \\
\midrule
\multirow{2}{*}{50} & $k$-prot & 0.29 (0.27) & 0.45 (0.26) & 0.56 (0.29) \\
                    & JM-mix & \textbf{0.69} (0.27) & \textbf{0.77} (0.26) & \textbf{0.89} (0.29) \\
\midrule
\multirow{2}{*}{100} & $k$-prot & 0.41 (0.19) & 0.58 (0.18) & 0.70 (0.21) \\
                     & JM-mix & \textbf{0.76} (0.19) & \textbf{0.88} (0.18) & \textbf{0.91} (0.21) \\
\midrule
\multirow{2}{*}{500} & $k$-prot & 0.39 (0.12) & 0.86 (0.14) & 0.94 (0.03) \\
                     & JM-mix & \textbf{0.92} (0.12) & \textbf{0.95} (0.06) & \textbf{0.97} (0.03) \\
\bottomrule
\toprule
\multicolumn{5}{c}{20\% Random Missing Data} \\
\midrule
$T$ & Method & $P=25$ & $P=50$ & $P=75$ \\
\midrule
\multirow{2}{*}{50} & $k$-prot & 0.21 (0.25) & 0.38 (0.26) & 0.46 (0.27) \\
                    & JM-mix & \textbf{0.68} (0.25) & \textbf{0.68} (0.26) & \textbf{0.83} (0.27) \\
\midrule
\multirow{2}{*}{100} & $k$-prot & 0.35 (0.20) & 0.48 (0.18) & 0.54 (0.16) \\
                     & JM-mix & \textbf{0.67} (0.20) & \textbf{0.80} (0.18) & \textbf{0.84} (0.16) \\
\midrule
\multirow{2}{*}{500} & $k$-prot & 0.36 (0.10) & 0.65 (0.11) & 0.75 (0.08) \\
                     & JM-mix & \textbf{0.77} (0.10) & \textbf{0.88} (0.11) & \textbf{0.87} (0.08) \\
\bottomrule
\end{tabular}
\end{table}


\section{Application to air quality data}
\label{sec:empstud}
We use the JM-mix to track air quality levels, treating it as a latent process that influences observable continuous and categorical variables. Our analysis focuses on air quality data from ARPA Lombardia\footnote{Data available here \url{https://www.arpalombardia.it/dati-e-indicatori/}.} for Milan, spanning June 5, 2021, to June 5, 2024. 

\subsection{Data Description}
Our dataset includes daily weather data on average, maximum, and minimum air temperature (Temp, °$C$), relative humidity (RH, \%), rainfall (RF, $mm$), global radiation (GR, $W/m^2$), and wind speed (WS, $m/s$). These measurements are taken from the weather station at Piazza Zavattari, Milan, which was selected because it provides data for all the aforementioned variables. 

Additionally, we collect air pollutant data on daily average, maximum, and minimum levels of nitrogen dioxide (NO$_2$) and ozone (O$_3$), as well as daily average levels of particulate matter with diameter around 2.5 $\mu m$ (PM$_{2.5}$), and with diameter around 10 $\mu m$ (PM$_{10}$).
The ARPA Lombardia air quality monitoring network consists of fixed stations that provide continuous data through automatic analyzers at regular intervals \citep{arpa2023stations,maranzano2024arpaldata}. This data includes aggregated municipal values from modeling simulations integrated with the monitoring network data.

\begin{figure}
    \centering
    \includegraphics[width=\linewidth]{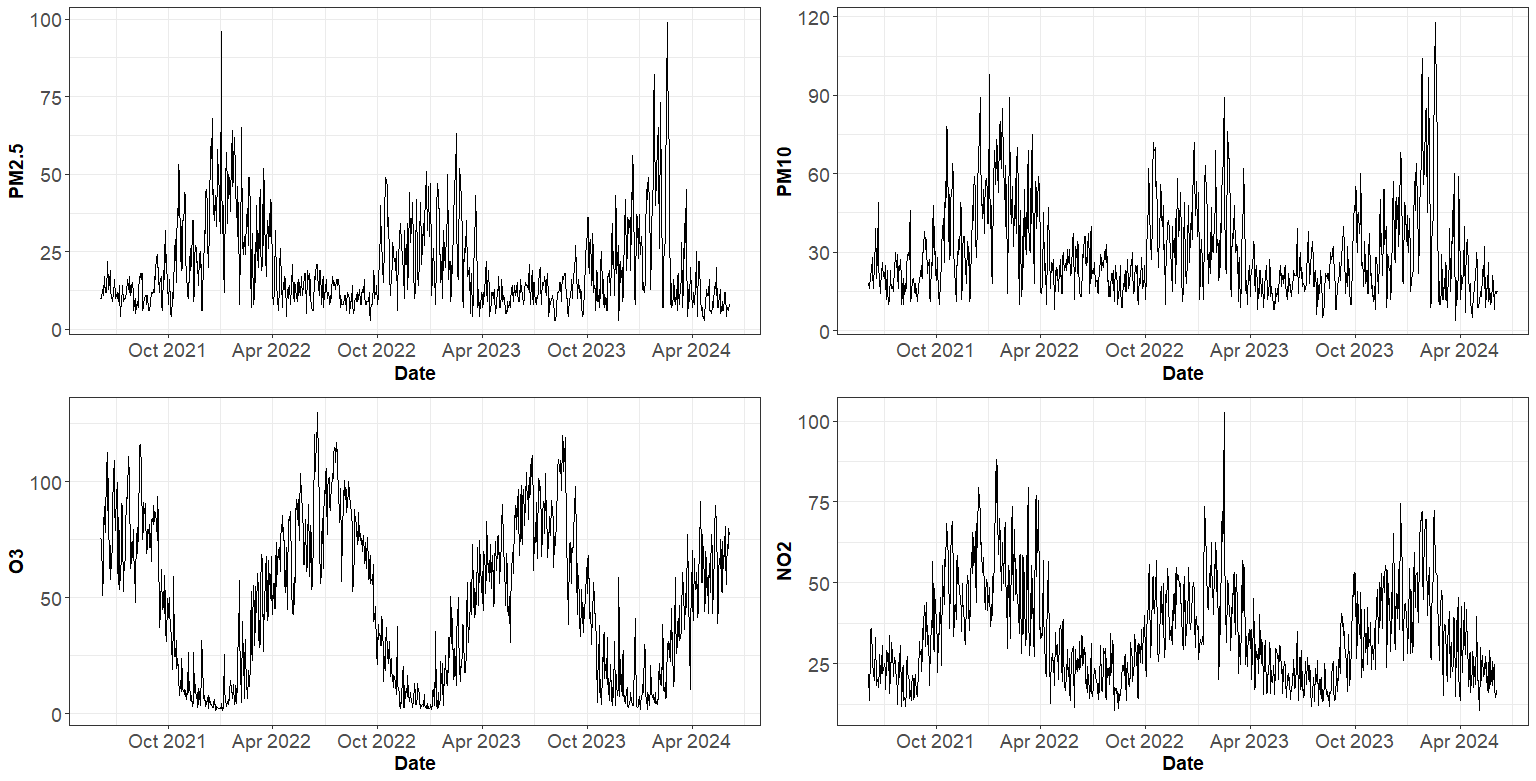}
    \caption{
    Daily time series of pollutants PM$_{2.5}$, PM$_{10}$, O$_3$, and NO$_2$ observed for the period June 2021 to June 2024.
    }
    \label{fig:pollut}
\end{figure}

\begin{figure}
    \centering
    \includegraphics[width=\linewidth]{ 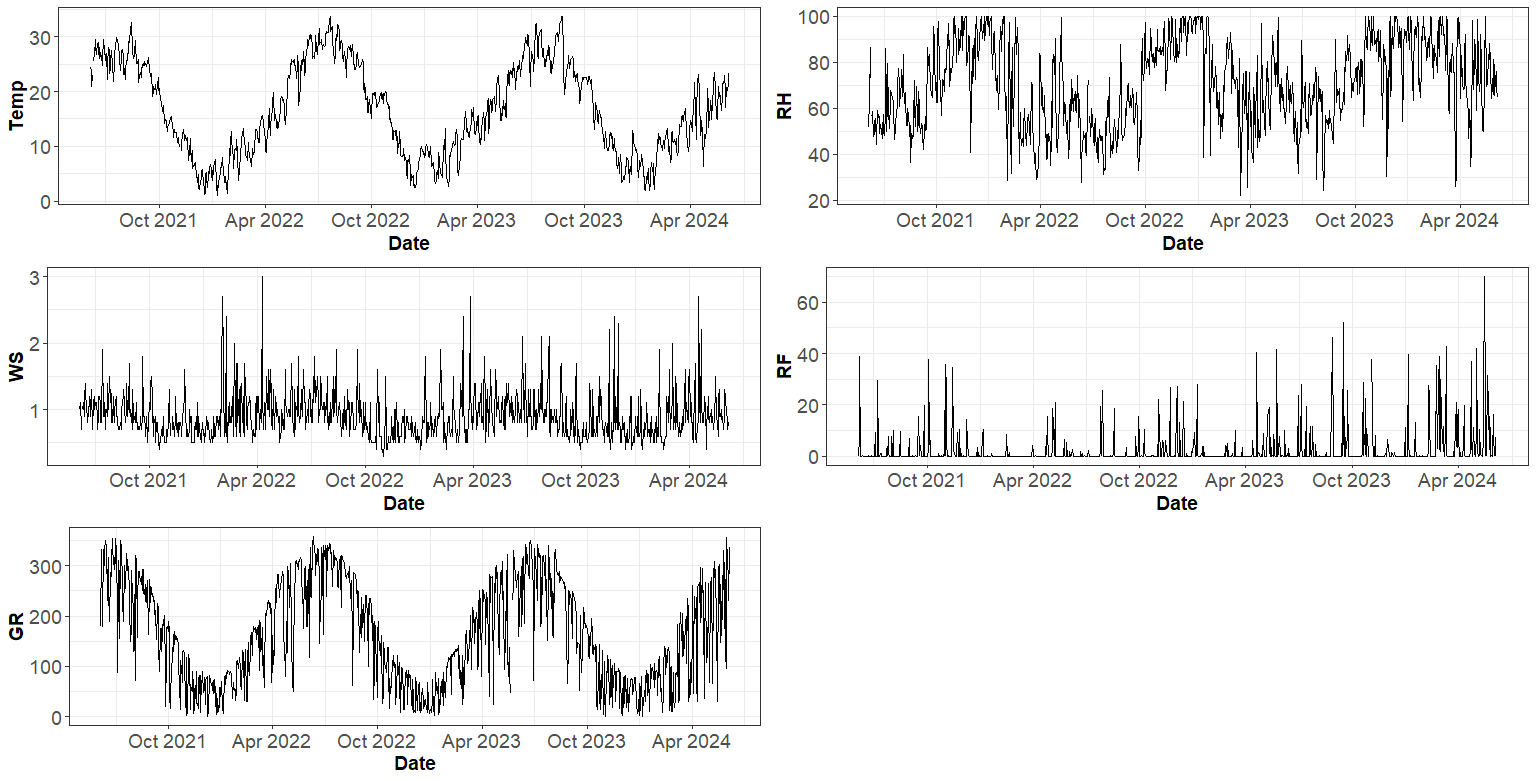}
    \caption{
    Daily time series temperature (Temp), relative humidity (RH), wind speed (WS), rainfall (RF), and global radiation (GB) observed for the period June 2021 to June 2024.
    }
    \label{fig:weath}
\end{figure}

Figures \ref{fig:pollut} and \ref{fig:weath} show the time series of pollutant and weather data. O$_3$ trends similarly to Temp and GR, while NO$_2$, PM$_{2.5}$, and PM$_{10}$ levels are higher in winter.

We incorporate categorical features such as wind speed greater\footnote{This threshold corresponds to `light air'' on the Beaufort scale.
} 
than 0.7 $m/s$, which is the observed median value (windy feature: ``Yes'' for windy days); similarly, we include a rainy feature (``Yes'' for rainy days) and consider a day to be rainy when rainfall is above the observed median value of 0 $mm$. Additional categorical features are month (1 to 12), weekend (yes or no), and holiday (yes or no).
The latter two variables are included to account for changes in human activity that can affect pollution levels.
Furthermore, we include 7-day moving averages of NO$_2$, O$_3$, PM$_{2.5}$, PM$_{10}$, Temp, WS, RF, RH, and GR. Additionally, similar as in \cite{cortese:2023}, we calculate 7-day moving correlations between each weather variable and each air pollutant. The moving correlation is computed by calculating the Pearson correlation coefficient over a rolling 7-day window, providing a dynamic view of the relationship between variables over time. 
The final dataset comprises 1,097 time observations and 56 features.

\subsection{Summary statistics}

Table \ref{table:summary_statistics} presents the summary statistics and percentage of missing values for continuous variables in the dataset. The table includes the mean, standard deviation (SD), minimum, and maximum values for each variable. The percentage of missing values (NA \%) indicates the proportion of missing data for each variable. The values reveal that Temp, RH, WS, RF and GR have a small percentage of missing data, while other variables have no missing values.
\begin{table}[h!]
    \centering
    \caption{
    Summary statistics and percentage of missing values for continuous variables.
    }
    \begin{tabular}{l|rrrrr}
        \hline
        Variable & Mean & SD & Min & Max & NA \% \\
        \hline
        PM$_{2.5}$ & 19.88 & 13.79 & 3.00 & 99.00 & 0.00 \\
        PM$_{10}$ & 30.28 & 17.30 & 4.00 & 118.00 & 0.00 \\
        O$_3$ & 48.18 & 32.60 & 1.40 & 129.80 & 0.00 \\
        NO$_2$ & 34.73 & 15.07 & 10.50 & 102.60 & 0.00 \\
        Temp & 16.70 & 8.28 & 1.00 & 33.70 & 0.36 \\
        RH & 70.91 & 18.42 & 22.00 & 100.00 & 0.36 \\
        WS & 0.90 & 0.34 & 0.30 & 3.00 & 0.36 \\
        RF & 2.36 & 6.95 & 0.00 & 70.00 & 0.36 \\
        GR & 162.85 & 100.56 & 0.40 & 357.20 & 0.27 \\
        \hline
    \end{tabular}
    \label{table:summary_statistics}
\end{table}

The mean values indicate that the average levels of PM$_{2.5}$ and PM$_{10}$ are below the standard safety limits \citep{world2021global}, but their maximum values show that there are instances of very high pollution. O$_3$ and NO$_2$ also show high variability, with maximum values indicating significant pollution episodes. Temp and RH exhibit substantial variability, reflecting seasonal changes. WS and RF have relatively low average values but show occasional extreme events. GR shows a wide range, indicating variability in solar radiation received.

Table \ref{table:correlation_matrix} displays the correlation matrix for continuous variables. Correlations greater than 0.5 or lower than -0.5 are highlighted in bold to emphasize strong relationships between variables. 
%
%
\begin{table}[h!]
    \centering
    \caption{
    Correlation (upper right) and partial correlation (lower left) matrix for continuous variables with correlations greater than 0.5 or lower than -0.5 highlighted in bold. Partial correlation coefficients are in italics.
    }
    \begin{tabular}{l|rrrrrrrrr}
        \hline
        \textbf{} & PM$_{2.5}$ & PM$_{10}$ & O$_3$ & NO$_2$ & Temp & RH & WS & RF & GR \\
        \hline
        PM$_{2.5}$ & - & \textbf{0.97} & \textbf{-0.58} & \textbf{0.76} & \textbf{-0.53} & 0.40 & -0.44 & -0.16 & -0.47 \\
        PM$_{10}$ & \textbf{\textit{0.93}} & - & \textbf{-0.51} & \textbf{0.75} & -0.43 & 0.31 & -0.43 & -0.20 & -0.39 \\
        O$_3$ & \textit{0.05} & \textit{0.01} & - & \textbf{-0.73} & \textbf{0.87} & \textbf{-0.74} & 0.39 & -0.06 & \textbf{0.86} \\
        NO$_2$ & \textit{-0.10} & \textit{0.33} & \textit{-0.39} & - & \textbf{-0.68} & 0.33 & -0.47 & -0.11 & \textbf{-0.56} \\
        Temp & \textit{-0.37} & \textit{0.39} & \textit{0.45} & \textit{-0.24} & - & \textbf{-0.59} & 0.20 & -0.07 & \textbf{0.80} \\
        RH & \textit{0.16} & \textit{-0.04} & \textit{-0.42} & \textbf{\textit{-0.50}} & \textit{0.11} & - & \textbf{-0.33} & 0.33 & \textbf{-0.79} \\
        WS & \textit{-0.09} & \textit{0.10} & \textit{0.11} & \textit{-0.35} & \textit{-0.29} & \textit{-0.34} & - & 0.20 & 0.23 \\
        RF & \textit{0.03} & \textit{-0.12} & \textit{0.23} & \textit{0.22} & \textit{0.07} & \textit{0.37} & \textit{0.25} & - & -0.25 \\
        GR & \textit{0.11} & \textit{-0.10} & \textit{0.29} & \textit{-0.06} & \textit{0.25} & \textit{-0.41} & \textit{-0.17} & \textit{-0.15} & - \\
        \hline
    \end{tabular}
    \label{table:correlation_matrix}
\end{table}
We observe strong positive relationships between PM$_{2.5}$ and PM$_{10}$, and strong negative relationships between O$_3$ and NO$_2$. Temperature is positively correlated with O$_3$ and GR, indicating higher ozone levels and global radiation during summer time. In contrast, temperature is negatively correlated with NO$_2$ and RH, suggesting lower nitrogen dioxide levels and humidity with higher temperatures.

Table \ref{table:frequency_table} shows the percentage of observations for each category and the percentage of missing values for each categorical variable. 
\begin{table}[h!]
    \centering
    \caption{
    Frequency table for the categorical variables.
    }
    \begin{tabular}{l|ccc}
        \hline
        Variable & No (\%) & Yes (\%) & NAs (\%) \\
        \hline
        Windy & 51.64 & 48.00 & 0.36 \\
        Rainy & 69.64 & 30.00 & 0.36 \\
        Weekend & 71.37 & 28.63 & 0 \\
        Holiday & 96.72 & 3.28 & 0 \\
        \hline
    \end{tabular}
    \label{table:frequency_table}
\end{table}
The percentages reveal that windy days are almost equally distributed, while there is a substantial prevalence of non-rainy days. Weekends account for about 29\% of the days, while holidays are rare, making up only about 3\% of the observations. 

\subsection{Results}
We select the model based on the modified version of the GIC by \cite{cortese2023b} and \cite{cortese2024generalized}, described in Section \ref{subsec:hyperpar}. 
According to this, we select $\lambda=0.30$. 
The GIC selects $K=2$, but we set $K=4$ following the air quality index \citep[AQI,][]{rule2005environmental} levels computed from observed pollutants (O$_3$, NO$_2$, PM$_{2.5}$, and PM$_{10}$), which correspond to four categories: ``Good'', ``Moderate'', ``Unhealthy for Sensitive Groups'' (US), and ``Unhealthy''. This choice of $K=4$ is a compromise between model fit and interpretability, aligning with the AQI categories for better practical relevance and ease of understanding.

Table \ref{table:state_means_modes} presents the conditional means and modes for the variables in each air quality state. This table allows us to comprehensively describe each state in terms of air quality, providing insights into the typical environmental conditions associated with each category. Based on these conditional means and modes, we assign meaningful labels to each state, using terminology from the AQI. For example, the ``Unhealthy'' state is characterized by higher levels of PM$_{2.5}$, PM$_{10}$, and NO$_2$, while the ``Good'' state corresponds to lower values of these pollutants.
Additionally, ``Good'' and ``Moderate'' air quality states are often associated with windy days, contrasting with the ``Unhealthy'' and ``US'' states, which tend to occur calmer days. December and October emerge as the most polluted months, suggesting seasonal influences on air quality. The absence of a significant difference in the rainy, weekend, and holiday features across states is likely due to the overall prevalence of non-rainy days and working days, as shown in Table \ref{table:frequency_table}. 
\begin{table}[h!]
    \centering
    \caption{
    State conditional means and modes.
    }
    \begin{tabular}{l|rrrr}
        \hline
        Variable & Good & Moderate & US & Unhealthy \\
        \hline
        PM$_{2.5}$ & 12.13 & 14.27 & 23.21 & 31.59 \\
        PM$_{10}$ & 21.67 & 23.53 & 36.01 & 42.23 \\
        O$_3$ & 82.87 & 58.29 & 29.81 & 10.57 \\
        NO$_2$ & 22.43 & 31.43 & 40.64 & 48.26 \\
        Temp & 26.42 & 15.38 & 14.89 & 7.14 \\
        RH & 57.48 & 64.04 & 80.89 & 85.84 \\
        WS & 1.00 & 1.01 & 0.77 & 0.77 \\
        RF & 1.69 & 3.25 & 3.51 & 1.66 \\
        GR & 260.02 & 191.92 & 109.53 & 58.70 \\
        Windy & Yes & Yes & No & No \\
        Rainy & No & No  & No  & No  \\
        Month & 7 & 4 & 10 & 12 \\
        Weekend & No & No  & No  & No \\
        Holiday & No & No & No & No  \\
        \hline
    \end{tabular}
    \label{table:state_means_modes}
\end{table}

Table \ref{table:visits_state} shows the percentage of visits in each state, reflecting how often the air quality falls into each of the four categories. The ``Good'' category has the highest percentage (32.63), followed by ``Unhealthy'' (26.62), ``Moderate'' (21.70), and US (19.05). 
\begin{table}[h!]
    \centering
    \caption{
    Percentage of visits in each state.
    }
    \begin{tabular}{l|c}
        \hline
        State & Percentage (\%) \\
        \hline
        Good & 32.63 \\
        Moderate & 21.70 \\
        US & 19.05 \\
        Unhealthy & 26.62 \\
        \hline
    \end{tabular}
    \label{table:visits_state}
\end{table}

Table \ref{table:state_correlations} details the correlation and partial correlation coefficients conditional on 
each state. This table helps to understand the relationships between different environmental factors under varying air quality conditions. For example, in the ``Good'' state, there is a negative correlation between PM$_{2.5}$ or PM$_{10}$ and Temp, while in the ``US'' state, this correlation approaches zero. These are coherent with their unconditional versions shown in Table \ref{table:correlation_matrix}. Additionally, the partial correlation between RF and PM$_{10}$ is negative in the US and Unhealthy states, while for PM$_{2.5}$ it is low. This observation indicates that rainfall might be more effective in removing larger particulate matter from the atmosphere compared to smaller particles.

Our model provides deeper insights into these correlation patterns, highlighting how they vary between different air quality states.

\begin{table}[h!]
    \centering
    \caption{
    State conditional correlation and partial correlation coefficients.
    }
    \begin{tabular}{l|rrrr|rrrr}
        \hline
          & \multicolumn{4}{c|}{Conditional Correlation} & \multicolumn{4}{c}{Conditional Partial Correlation} \\
        Variable & Good & Moderate & US & Unhealthy & Good & Moderate & US & Unhealthy \\
        \hline
        PM$_{2.5}$, Temp & -0.33 & -0.30 & -0.12 & -0.33 & \textit{0.01} & \textit{-0.22} & \textbf{\textit{-0.59}} & \textit{-0.36} \\
        PM$_{2.5}$, RH & 0.29 & -0.20 & -0.10 & 0.29 & \textit{0.33} & \textit{0.10} & \textit{0.21} & \textit{-0.08} \\
        PM$_{2.5}$, WS & -0.38 & -0.24 & -0.46 & -0.38 & \textit{-0.10} & \textit{-0.11} & \textit{0.03} & \textit{0.01} \\
        PM$_{2.5}$, RF & -0.18 & -0.18 & -0.29 & -0.18 & \textit{-0.03} & \textit{-0.06} & \textit{0.15} & \textit{0.16} \\
        PM$_{2.5}$, GR & 0.06 & -0.17 & 0.10 & 0.06 & \textit{0.04} & \textit{-0.07} & \textit{0.21} & \textit{0.02} \\
        PM$_{10}$, Temp & -0.26 & -0.23 & 0.01 & -0.26 & \textit{0.20} & \textit{0.20} & \textbf{\textit{0.60}} & \textit{0.37} \\
        PM$_{10}$, RH & 0.19 & -0.25 & -0.14 & 0.19 & \textit{-0.11} & \textit{-0.01} & \textit{-0.16} & \textit{0.19} \\
        PM$_{10}$, WS & -0.35 & -0.22 & \textbf{-0.52} & -0.35 & \textit{0.13} & \textit{0.15} & \textit{-0.06} & \textit{0.03} \\
        PM$_{10}$, RF & -0.23 & -0.20 & -0.36 & -0.23 & \textit{-0.08} & \textit{-0.03} & \textit{-0.22} & \textit{-0.21} \\
        PM$_{10}$, GR & 0.16 & -0.11 & 0.15 & 0.16 & \textit{-0.13} & \textit{0.03} & \textit{-0.23} & \textit{0.03} \\
        O$_3$, Temp & 0.33 & 0.49 & 0.39 & 0.33 & \textit{0.34} & \textit{0.33} & \textit{0.28} & \textit{-0.06} \\
        O$_3$, RH & \textbf{-0.72} & -0.31 & -0.31 & \textbf{-0.72} & \textit{-0.36} & \textbf{\textit{-0.52}} & \textit{-0.29} & \textbf{\textit{-0.53}} \\
        O$_3$, WS & \textbf{0.74} & 0.21 & 0.28 & \textbf{0.74} & \textit{0.07} & \textit{-0.03} & \textit{0.28} & \textit{0.45} \\
        O$_3$, RF & -0.02 & -0.05 & -0.05 & -0.02 & \textit{0.11} & \textit{0.28} & \textit{0.11} & \textit{0.16} \\
        O$_3$, GR & 0.46 &\textbf{ 0.54} & 0.37 & 0.46 & \textit{0.14} & \textit{0.11} & \textit{0.25} & \textit{0.29} \\
        NO$_2$, Temp & -0.32 & -0.38 & -0.24 & -0.32 & \textit{-0.22} & \textit{0.02} & \textit{-0.24} & \textit{-0.42} \\
        NO$_2$, RH & -0.18 & -0.31 & -0.27 & -0.18 & \textit{-0.45} & \textbf{\textit{-0.57}} & \textit{-0.32} & \textbf{\textit{-0.61}} \\
        NO$_2$, WS & -0.34 & -0.34 & \textbf{-0.54} & -0.34 & \textit{-0.28} & \textit{-0.37} & \textit{-0.23} & \textit{-0.20} \\
        NO$_2$, RF & -0.20 & -0.20 & -0.25 & -0.20 & \textit{0.17} & \textit{0.24} & \textit{0.24} & \textit{0.29} \\
        NO$_2$, GR & 0.29 & -0.15 & 0.17 & 0.29 & \textit{-0.07} & \textit{-0.10} & \textit{0.08} & \textit{0.14} \\
        \hline
    \end{tabular}
    \label{table:state_correlations}
\end{table}

\begin{figure}
    \centering
    \includegraphics[width=\linewidth]{ 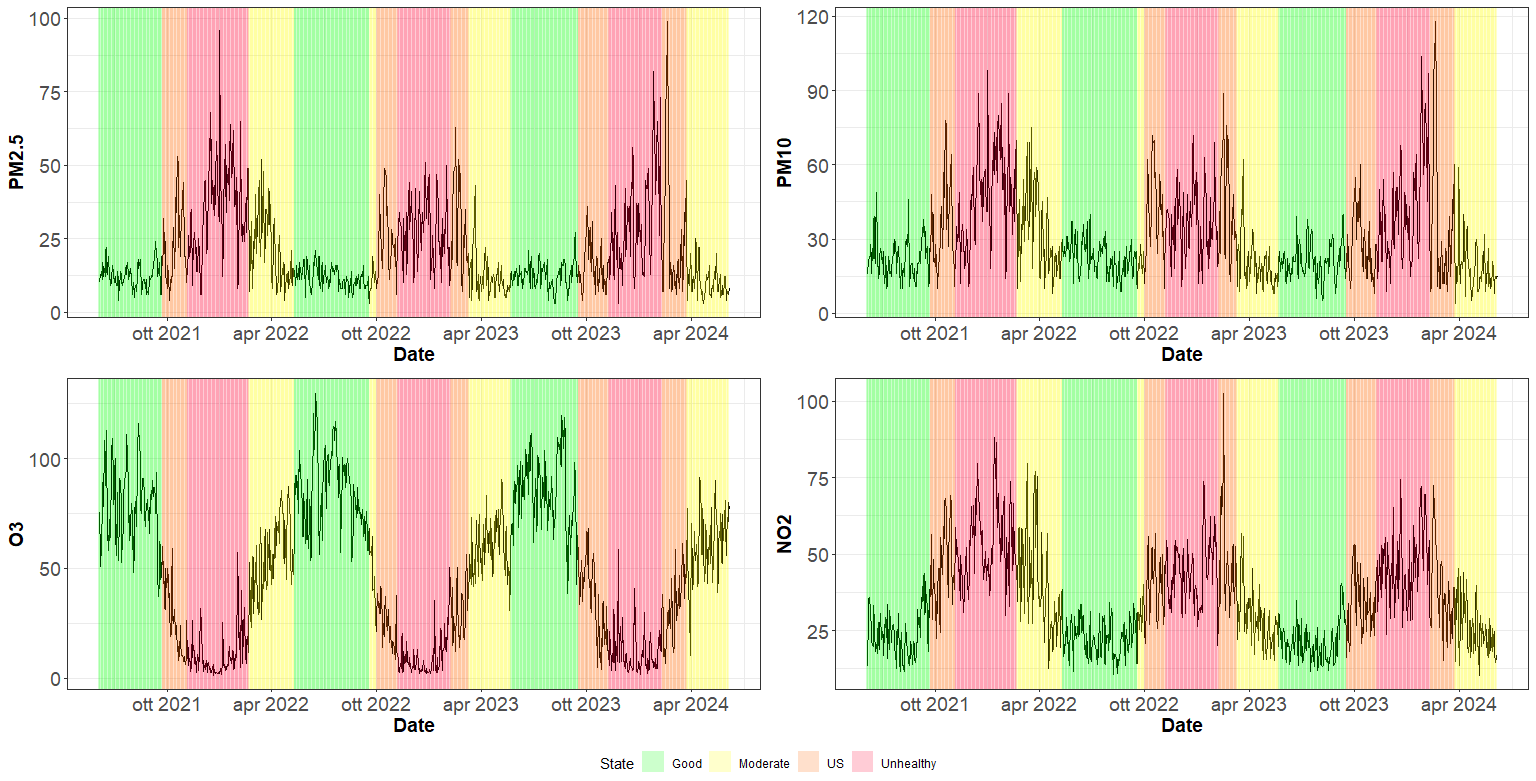}
    \caption{
    Daily time series of pollutants PM$_{2.5}$, PM$_{10}$, O$_3$, and NO$_2$ observed for the period June 2021 to June 2024, with the decoded state sequence fitted by the JM-mix. Colors represent different air quality states: green is ``Good``, yellow is ``Moderate``, orange is ``Unhealthy for Sensitive Groups (US)``, and red is ``Unhealthy``.
    }
    \label{fig:pollut_state}
\end{figure}

\begin{figure}
    \centering
    \includegraphics[width=\linewidth]{ 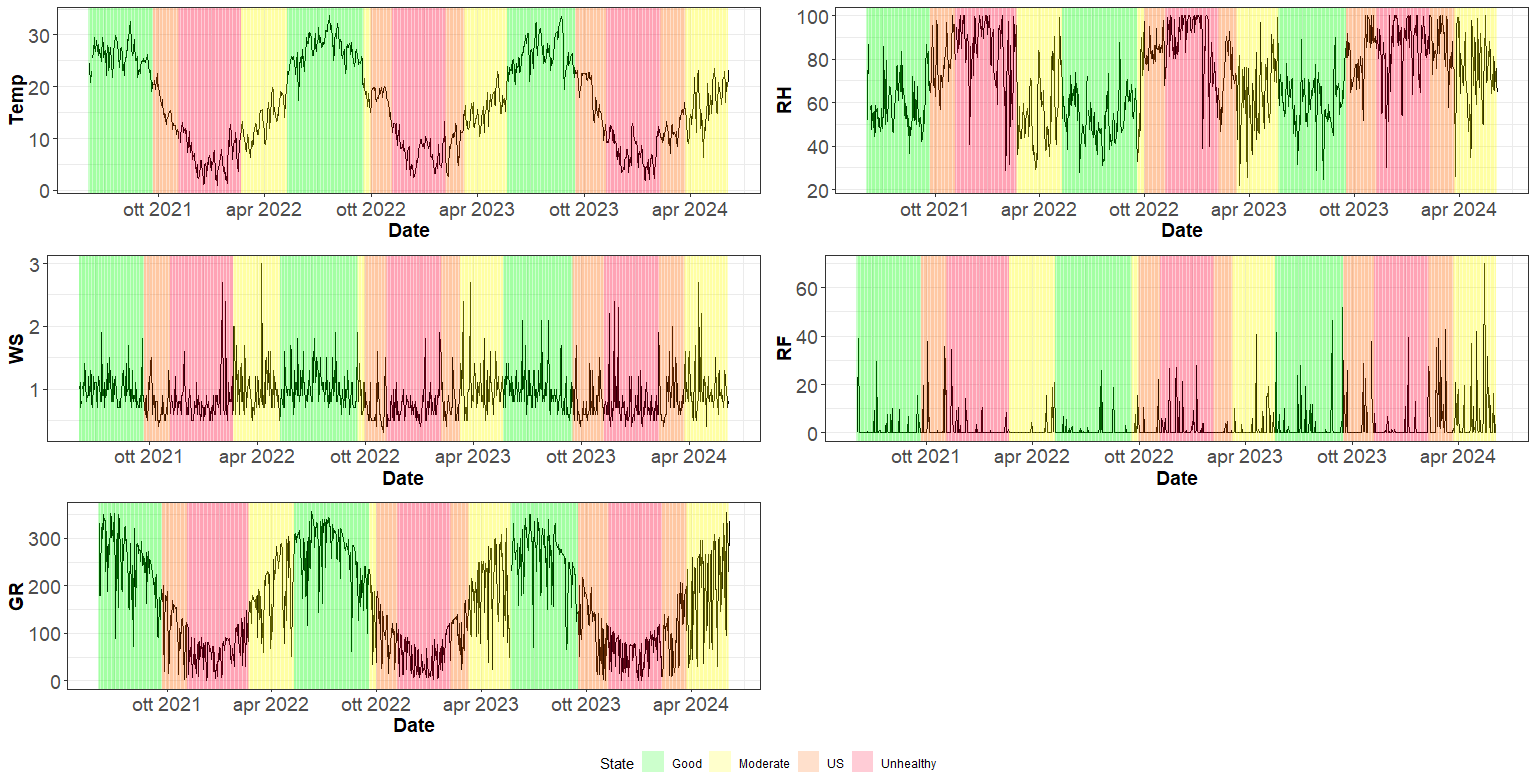}
    \caption{
    Daily time series temperature (Temp), relative humidity (RH), wind speed (WS), rainfall (RF), and global radiation (GB) observed for the period June 2021 to June 2024, with the decoded state sequence fitted by the JM-mix. Colors represent different air quality states: green is ``Good``, yellow is ``Moderate``, orange is ``Unhealthy for Sensitive Groups (US)``, and red is ``Unhealthy``.
    }
    \label{fig:weath_state}
\end{figure}

Figures \ref{fig:pollut_state} and \ref{fig:weath_state} illustrate the temporal variations of air pollutants and meteorological variables including background color bands representing air quality categories: ``Good'' (green), ``Moderate'' (yellow), ``US'' (orange), and ``Unhealthy'' (red).
We observe significant seasonal variations in PM$_{2.5}$ and PM$_{10}$, with higher concentrations during colder months, likely due to increased heating activities and stable atmospheric conditions \citep{sharma2021seasonal}. Peaks frequently enter the ``Unhealthy'' range. O$_3$ levels exhibit a distinct seasonal cycle, with higher concentrations during warmer months due to increased photochemical activity, frequently fluctuating between ``Good'' and ``Moderate'' categories. NO$_2$ levels display seasonal fluctuations with higher values during colder months, possibly due to increased emissions from heating and reduced atmospheric dispersion, sometimes reaching the US category.

\begin{figure}
    \centering
    \includegraphics[width=\linewidth]{ 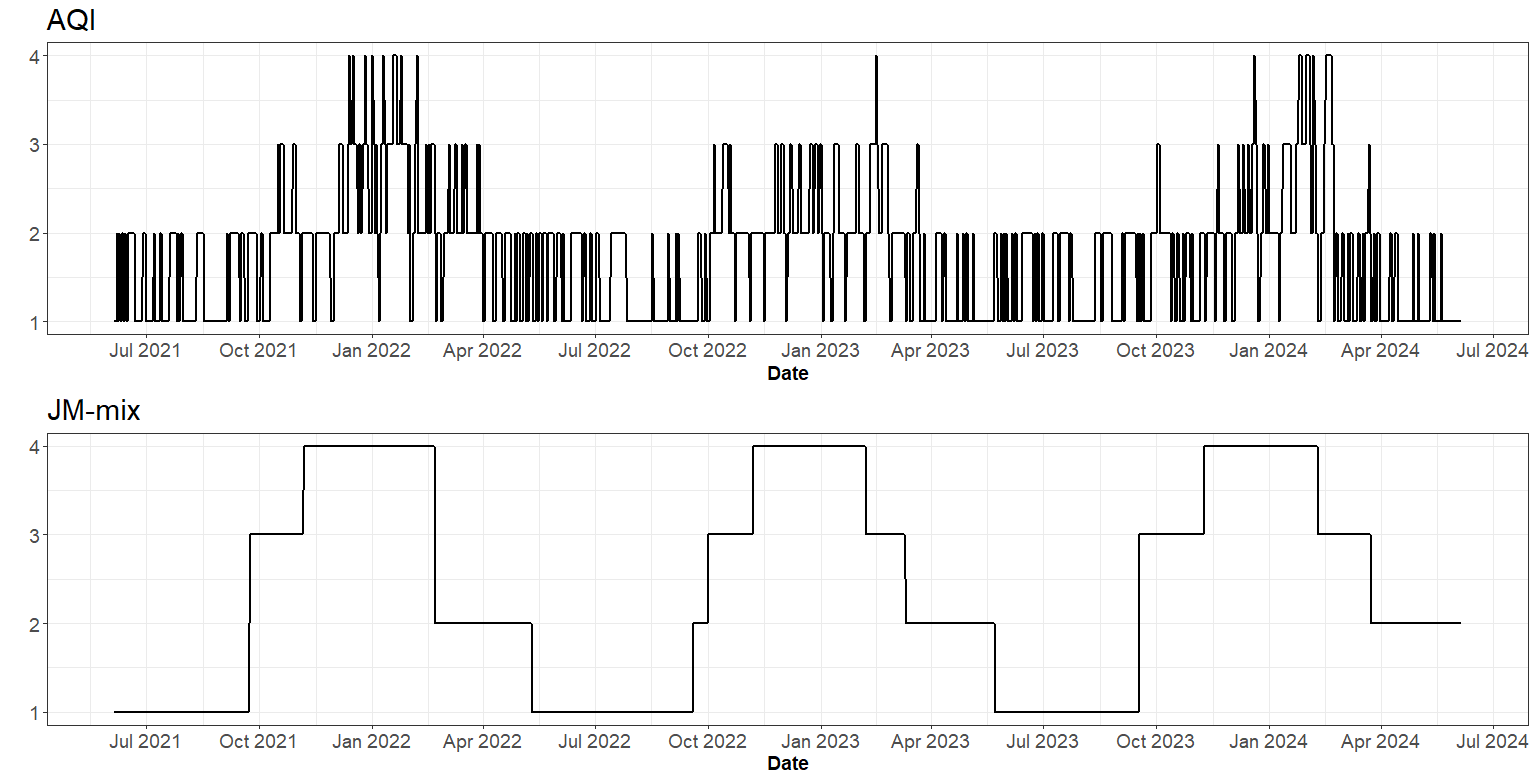}
    \caption{
    Visual comparison between the Air Quality Index (AQI) and latent states of the JM-mix. The states are ordered as follows: from 1 to 4, ``Good'', ``Moderate'', ``Unhealthy for Sensitive Groups'', and ``Unhealthy''.
    }
    \label{fig:AQIvsJMmix}
\end{figure}

Figure \ref{fig:AQIvsJMmix} shows the time evolution of the state sequences fitted with JM-mix and the AQI computed with the standard formula. The AQI for pollutant $q$, denoted as AQI$_q$, is given by
\[ \text{AQI}_q = \frac{(I_{high} - I_{low})}{(C_{high} - C_{low})} \times (C_q - C_{low}) + I_{low}, \]
where \( \text{AQI}_q \) is the AQI for pollutant \( q \), \( C_q \) is the concentration of pollutant \( q \), \( C_{low} \) and \( C_{high} \) are the breakpoints closest to \( C_q \), and \( I_{low} \) and \( I_{high} \) are the AQI values corresponding to \( C_{low} \) and \( C_{high} \). 
The general AQI is just the maximum AQI$_q$ among all pollutants.

The AQI sequence appears highly volatile, indicating rapid shifts in air quality due to its sensitivity to short-term fluctuations in pollutant concentrations. In contrast, the JM-mix sequence shows more stable transitions between states, suggesting it captures more consistent and prolonged periods of air quality. Essentially, the JM-mix acts as a smoothing mechanism for the categorical AQI variable, providing a clearer and more stable representation of air quality over time.

The differences between the AQI and JM-mix sequences arise from their distinct nature. The AQI is computed using a standard formula that primarily considers the immediate concentrations of key pollutants (O$_3$, NO$_2$, PM$_{2.5}$, and PM$_{10}$). It reacts quickly to changes in these pollutants, hence the high volatility. 
The JM-mix incorporates a broader set of variables: this comprehensive approach allows it to provide a more nuanced representation of air quality states, considering the underlying factors that influence pollutant levels.

The reduced volatility in the JM-mix sequence might be beneficial in contexts where stability and consistency are crucial. For instance, in public health and policy-making, having a stable representation of air quality can lead to better short to medium term planning and responses. 

\section{Discussion}
\label{sec:discussion}

In this work, we make three main contributions. First, we introduce a statistical jump model for mixed-type data, a novel method for clustering mixed-type data over time, incorporating regime persistence to enhance interpretability and utility. Second, our approach provides a fast and efficient solution for managing missing data, ensuring robust clustering results. Third, we demonstrate the applicability of our method through an empirical study on air quality data, showing that the proposal can infer more persistent air quality regimes compared to the traditional air quality index.

Simulation studies show that our model outperforms existing models like spectral clustering and $k$-prototypes in terms of clustering accuracy and robustness against missing data. 

In future work, we aim to investigate the issue raised by \cite{hennig2013find}, who argue that categorical variables may dominate distance-based clustering unless weighted down. This is significant for practical applications, and our proposed algorithm allows for setting variable-specific weights for clustering. A potential variable selection scheme could be based on \cite{hennig2023quantifying}, who introduced variable importance to assess the influence of each variable on clustering results. Another approach could involve a sparse version of our jump model with mixed-type data, as in \cite{nystrup:2021}, using Gower distance weights as defined in Appendix \ref{appendix A} and applying soft thresholding for weights estimation.

\section*{Declarations}

\begin{itemize}
\item[] \textbf{Funding:} 
This work has been supported by the European Commission, NextGenerationEU, Mission 4 Component 2, ``Dalla ricerca all’impresa'', Innovation Ecosystem RAISE ``Robotics and AI for Socio-economic Empowerment'', ECS00000035.
\item[] \textbf{Conflict of interest/Competing interests:} The authors declare that there are no conflicts of interest.
\item[] \textbf{Data availability:} Data is available upon request from the corresponding author.
\item[] \textbf{Code availability:} Code will be made available on the corresponding author's GitHub webpage.
\item[] \textbf{Author contribution:} 
Corresponding author contributed to the study conception and design. 
Second author supervised the entire project.
All authors contributed to material preparation, data collection and analysis. 
The first draft of the manuscript was written by corresponding author and all authors commented on previous versions of the manuscript. All authors read and approved the final manuscript.
\end{itemize}

%
%
%
%
%
%
\clearpage

\begin{appendices}

\section{ }
\label{appendix A}

The Gower distance \citep{gower1971general} between two vectors $\boldsymbol{x}_{i,\cdot}, \boldsymbol{x}_{j,\cdot} \in \mathbb{R}^P$, is defined as:

\begin{equation}
\label{eq:gower}
    g(\boldsymbol{x}_{i,\cdot}, \boldsymbol{x}_{j,\cdot}) = \frac{\sum_{p=1}^{P} w_{ijp} d_{ijp}}{\sum_{p=1}^{P} w_{ijp}}
\end{equation}

where:

\begin{itemize}
    \item $d_{ijp}$ is the contribution of feature $p=1,\ldots,P$ to the distance between $\boldsymbol{x}_{i,\cdot}$ and $\boldsymbol{x}_{j,\cdot}$.
    \item $w_{ijp}$ is a weight assigned to feature $p$ for the pair $(\boldsymbol{x}_{i\cdot}, \boldsymbol{x}_{j\cdot})$. We assume that all these weights are equal to 1.
\end{itemize}

The specific contribution $d_{ijp}$ depends on the type of feature $p$:

\begin{itemize}
    \item For continuous features: 
    \[
    d_{ijp} = \frac{|x_{ip} - x_{jp}|}{\max(\boldsymbol{x}_{\cdot, p}) - \min(\boldsymbol{x}_{\cdot, p})}
    \]
    where $\max(\boldsymbol{x}_{\cdot, p})$ and $\min(\boldsymbol{x}_{\cdot, p})$ are the maximum and minimum values of feature $p$ across all data points $\boldsymbol{x}_{\cdot, p}=(x_{1,p},\ldots, x_{T,p})\in \mathbb{R}^T$.

    \item For categorical features:
    \[
    d_{ijp} = 
    \begin{cases} 
    0 & \text{if } x_{ip} = x_{jp} \\
    1 & \text{if } x_{ip} \neq x_{jp}
    \end{cases}
    \]
\end{itemize}

\section{ }
\label{appendix B}
The adjusted Rand index \citep[ARI,][]{hubert:1985} measures the degree of overlap between two groupings of partitions: $G^m=\{G_1^m,\ldots,G_h^m\}$, obtained with a clustering algorithm on a set of $n$ elements, and $G^r=\{G_1^r,\ldots,G_q^r\}$, interpreted as the real clustering. 

Given the following contingency table:
\begin{equation}
\begin{array}{c|cccc|c}
\vspace{1em}
 & G_1^m & G_2^m & \cdots & G_h^m & \sum_{j=1}^h n_{ij} \\
\hline G_1^r & n_{11} & n_{12} & \cdots & n_{1 h} & a_1 \\
G_2^r & n_{21} & n_{22} & \cdots & n_{2 h} & a_2 \\
\vdots & \vdots & \vdots & \ddots & \vdots & \vdots \\
G_q^r & n_{q 1} & n_{q 2} & \cdots & n_{q h} & a_q \\
\hline \sum_{i=1}^q n_{ij}& b_1 & b_2 & \cdots & b_h & n
\vspace{1em}
\end{array}
\end{equation}
$\text{ARI}(G^r,G^m)$ can be calculated using the methodology proposed by \citet{ARIdef2} as
\begin{equation}
\label{eq:ARIeq}
\operatorname{ARI}\left(G^r, G^m\right) :=
\frac{\sum_{i j}
\binom{n_{ij}}{2}
-\frac{\left[\sum_i
\binom{a_{i}}{2}
\sum_j
\binom{b_{j}}{2}
\right]}{
\binom{n}{2}
}}{\frac{1}{2}\left[\sum_i
\binom{a_{i}}{2}
+\sum_j
\binom{b_{j}}{2}
\right]-\frac{\left[\sum_i
\binom{a_{i}}{2}
\sum_j
\binom{b_{j}}{2}
\right]}{
\binom{n}{2}
}} \,.
\end{equation}
Here, $n_{ij}$ represents the count of shared elements within clusters $G_i^m$ and $G_j^r$, while $a_i$ aggregates all $n_{ij}$ values linked to any $G_j^r$ in $G^r$, and all $G_i^m$ in $G^m$. Similarly, $b_j$ sums all $n_{ij}$ values related to any $G_i^m$ in $G^m$, and all $G_j^r$ in $G^r$.




\end{appendices}


\clearpage
\printbibliography

\end{document}